\documentclass[letterpaper,12pt]{article}
\usepackage[usenames,dvipsnames]{xcolor}
\usepackage{mathtools} 
\usepackage{jheppub2} 
\usepackage[T1]{fontenc}
\usepackage{natbib}
\usepackage{graphicx}
\usepackage{hyperref}
\usepackage{amsmath}
\usepackage{amsfonts}
\usepackage{amssymb}
\usepackage{enumitem}
\usepackage{tocvsec2}
\usepackage{slashed}
\usepackage{comment}
\usepackage{url}
\usepackage{pifont}
\usepackage{makecell}
\newcommand{\xmark}{\ding{55}}

\newcommand{\be}{\begin{equation}}
\newcommand{\ee}{\end{equation}}
\newcommand\beq{\begin{eqnarray}}
\newcommand\eeq{\end{eqnarray}}

\newcommand{\mybar}[1]

\hypersetup{
    colorlinks=true,   
    linkcolor=Cerulean,     
    citecolor=OrangeRed,    
    filecolor=magenta, 
    urlcolor=blue      
}

\makeatletter
\def\l@subsubsection#1#2{}
\makeatother    

\advance\footnotesep 1.5pt

\begin{document}

\title{Chiral fermions on lattice axion strings}
\author{Srimoyee Sen}
\emailAdd{srimoyee08@gmail.com}
\affiliation{
Department of Physics and Astronomy, Iowa State University, Ames, IA  50010
}

\abstract{
I discretize axion string configuration coupled to a Dirac fermion, which in the continuum binds a massless chiral fermion in its core when the winding is one. I show that such a configuration can host one or more chiral fermions when regulated on the lattice. Realization of these chiral fermions relies on the presence of Wilson-like terms similar to the Wilson term used in lattice domain wall fermions. The number of chiral fermions on the string jumps as the Wilson-like parameter is varied with respect to the other mass scales in the problem. These jumps coincide with phase transitions along a two dimensional surface passing through the string's core. A one-loop Feynman diagram is used to demonstrate how anomaly inflow works in this lattice regularized theory.}

\maketitle
\maxtocdepth{subsection} 
\section{Introduction}
In \cite{Callan:1984sa}, Callan and Harvey demonstrated that fermions coupled to a domain wall in odd number of dimensions and vortex defects in even dimensions ($2n+2$) can exhibit massless chiral edge states bound to the defects at low energy. For example, the low energy spectrum of a massive Dirac fermion in $2+1$ or $4+1$ dimensions, with a domain wall profile in its mass has a single chiral fermion bound to the domain wall. Similarly, in the even dimensional case $2n+2$, a Dirac fermion coupled to an axion string exhibits chiral edge states bound to the core of the string. These systems are of interest to both high energy and condensed matter physics. For instance, cosmic axion strings are considered to be important for astrophysical signature of new physics at high energy \cite{Agrawal:2020euj, Agrawal:2019lkr, Gorghetto:2018myk, Gorghetto:2020qws, Buschmann:2019icd}. They are also of relevance to the physics of early universe. 
Similarly, domain wall theories of fermions in odd space-time dimensions and axion strings coupled to fermions in even dimensions have analogs in condensed matter physics. The former describes the physics of quantum Hall effect
\cite{PhysRevLett.45.494, PhysRevLett.49.405, PhysRevLett.54.259, PhysRevB.31.3372} where as the latter can be realized in axion insulators \cite{Sekine:2020ixs, Varnava_2018} as discussed in \cite{Wang:2012bgb, Qi:2012cs, Schuster:2016ong}. The chiral edge states living on the defects in both cases suffer from chiral anomaly when fermion number symmetry is gauged and the defect theory by itself violates current conservation. Callan and Harvey showed that current conservation is restored by an inflowing current from the bulk to the defect. This inflowing current can be computed by integrating out the fermion sufficiently far away from the defect where the fermion spectrum is completely gapped. This results in an effective field theory for the gauge field which includes a Chern-Simons term in both cases. The Chern-Simons level is obtained by a one-loop Feynman diagram following the computation of Goldstone-Wilczek \cite{PhysRevLett.47.986}. Once this Chern-Simons term is obtained the inflow current can be computed by extremizing the Chern-Simons gauge field action. More generally, gapped fermion field theories in various space-time dimensions correspond to the physics of symmetry protected topological phases and noninteracting topological insulators in condensed matter physics \cite{Witten:2015aba, Ludwig_2015, PhysRevB.85.085103, Ryu_2010, Kitaev:2009mg, PhysRevB.78.195125, PhysRevB.90.245120, Wang:2022ucy}. The massless fermion edge states in these theories may or may not be chiral. As a result there may not be any inflowing Hall current from the bulk to the boundary. However, the bulk physics in these theories is often related to the the boundary states via global anomalies. More recently, the idea of Hall current has been generalized to include fermion theories with or without continuous symmetries in any number of dimensions where the edge states are not necessarily chiral \cite{Kaplan:2021ewi, longpaper}. 

The continuum construction of chiral fermion edge states on the domain wall was later used by \cite{Kaplan:1992bt} to realize chiral fermions on the lattice. This construction of lattice chiral fermions has the advantage of retaining global chiral symmetry and 
 has therefore been extensively used in simulations of QCD where this is a desirable feature. Interestingly, the construction of lattice domain wall fermion is more subtle than the continuum construction in Callan-Harvey. The subtleties arise due to the presence of fermion doublers coming from naive discretization of fermions. With naive discretization, the domain wall carries equal numbers of right and left moving doublers which preclude any net chirality on the wall. In order to eliminate the unwanted doublers, one has to use a Wilson-Dirac Lagrangian in a domain wall background for the mass term \cite{Kaplan:1992bt}. For certain values of the Wilson parameter $R$ and the Dirac mass $M$ one can realize a net imbalance of right and left moving modes on the wall resulting in a net chirality. The corresponding bulk theory away from the domain wall exhibits a topological phase analogous to those observed in Chern insulators in condensed matter physics \cite{Kaplan:1992bt, Jansen:1992yj, Golterman:1992ub, Jansen:1992tw, Jansen:1992tw}. The Wilson parameter in lattice QFT corresponds to the hopping parameter in Chern insulators and the Dirac mass corresponds to magnetic polarization\cite{PhysRevB.78.195424, Bernevig_2006}. These examples, being demonstrative of the deep ties between lattice fermion field theory and the physics of topological materials have inspired several papers in recent times \cite{Tirrito:2022vmk, Ziegler:2021yua, Ziegler:2020zkq, Sen:2020srn, Kaplan:2019pdd}.
 Given the subtleties associated with realization of chiral domain wall fermions and the rich phase diagram that accompanies this construction, it is interesting to ask how a lattice discretization of axion strings will alter the chiral fermion spectrum confined to the string.

Unsurprisingly, it is not just the lattice theory of domain wall fermions where the presence of fermion doublers eliminates any net chirality on the wall, naive discretization of the axion string theory behaves the same way, i.e. there are equal number of right and left moving doublers on the string. As a result the string carries no net chirality. As I outline in the next few sections of this paper, in order to realize a net chirality on the defect one must introduce Wilson-like terms in the Lagrangian to create an imbalance of right and left moving modes on the string. Additionally, the number of chiral fermions on the defect can change abruptly as a function of the parameters in the theory. 
Before discretizing space-time it is convenient to write the axion string in terms of a crossed domain wall configuration as illustrated in the next section. This configuration is easier to discretize and the corresponding Wilson-like terms have a simple form. A Similar crossed domain wall configuration in the continuum was studied in the context of regulating four dimensional chiral gauge theories in \cite{Fukaya:2016ofi}.

As stated earlier, in Callan-Harvey, the fermion number current which flows from the bulk to the boundary in a background electric field (Goldstone-Wilczek current),can be computed using a one-loop Feynman diagram. It was shown \cite{Golterman:1992ub} that a similar calculation applies for lattice domain wall fermions in the continuum limit, where the Goldstone-Wilczek current is computed using the same Feynman diagram as in the continuum, using lattice perturbation theory. The Feynman diagram on the lattice computes the winding number of a map from momentum space, which on the lattice is a torus, to the Dirac space, which is a sphere.  The result therefore is quantized. The winding number of this map jumps as a function of the Dirac mass $M$ and the Wilson parameter $R$ when the bulk fermion propagator goes gapless. As I show in this paper, a similar calculation applies to the axion string. The corresponding Goldstone-Wilczek current is computed using a one-loop Feynman diagram and the net current exhibits discrete jumps as a function of the parameters of the theory just as in lattice domain wall. These discrete changes in the Goldstone-WIlczek current are necessary to compensate for the boundary current as the number and chirality of edge states jump. Furthermore, I find that at certain values of the Wilson-like parameter, the bulk fermion gap goes to zero along a two dimensional surface passing through the defect, coinciding with the discrete jumps in the chiral edge states. This indicates that the discrete jump in chiral edge states is accompanied by a phase transition along this two dimensional surface.  

The organization of this paper is as follows. I begin with a brief review of the lattice construction of domain wall fermions which is followed by axion string analysis in the continuum. The subsequent section demonstrates that the axion string configuration is equivalent to a crossed domain wall configuration which is then discretized. The corresponding Wilson-like terms are introduced to engineer chiral edge states on the string and the associated Goldstone-Wilczek current is computed. This is followed by a section which discusses possible numerical realizations of this construction and its relevance to axion insulators. 

\section{Domain wall and vortex string}
To review the lattice domain wall construction, it is convenient to first focus on $2+1$ dimensions ($x_0, x_1, x_2$) where a Dirac fermion with a mass defect $m=m_0\epsilon(x_2)$ for $m_0>0$ exhibits chiral edge states on the defect (domain wall) at $x_2=0$. 
If this fermion theory is discretized naively, the low energy spectrum on the wall will carry equal number of right moving and left moving edge states. This is caused by fermion doubling as can be seen from the discretized Euclidean Dirac equation below
\beq
\left(i\gamma^{\mu}\frac{\sin(p_{\mu}a)}{a}+\gamma_2\nabla_2+m\right)\psi(p,x_2)=0
\label{disc}
\eeq
where $\mu$ takes values $0$ and $1$, the $x_0, x_1$ coordinates have been Fourier transformed and $\nabla_2$ is the lattice derivative $\frac{\delta_{x, x+a_2}-\delta_{x, x-a_2}}{2a}$ when the lattice spacing is $a$. Since I am interested in massless chiral edge states on the wall, I can set $\sum_{\mu}\gamma_{\mu}\sin(p_\mu a)=0$. 
As seen from this equation, the transverse profile for the states located near the corners of the Brillouin zones $p_0=n \frac{\pi}{a}, p_1=m\frac{\pi}{a}$ with $n, m=0,1$ are identical and all of these modes have normalizable transverse profiles. The modes around the BZ corners $\{0,0\}$ and $\{\frac{\pi}{a},\frac{\pi}{a}\}$ are of chirality $-1$ whereas $\{\frac{\pi}{a},0\}$ and $\{0,\frac{\pi}{a}\}$ are of chirality $+1$, thus eliminating any net chirality on the wall. 
It was shown in \cite{Kaplan:1992bt} that one needs to introduce a Wilson term, $\frac{R}{2}\bar{\psi}\nabla^2\psi$ in the Lagrangian in order to realize an imbalance between right and left moving modes. To see how this comes about, one can set $R=a$. Then the equation for the transverse profile for the edge states becomes
\beq
\psi(x_2+a)=-(m_\text{eff})\psi(x_2)
\eeq
where $m_{\text{eff}}=m a-1-F(p)$ with $F(p)=\sum_{\mu=0,1}(1-\cos(p_{\mu}a))$. This equation is solved by the ansatz $\psi=(-m_{\text{eff}})^{x_2/a}$ and a normalizable mode exists as long as 
\beq
2>m_0a -F(p)>0. 
\eeq
For $0<m_0 a<2$, it is only the states around $\{0,0\}$ which are normalizable. For $2<m_0 a<4$, the normalizable states are centered around $\{0, \frac{\pi}{a}\}$ and $\{\frac{\pi}{a},0\}$. Similarly for $4<m_0 a<6$, it's the states centered around $\{\frac{\pi}{a},\frac{\pi}{a}\}$ which have normalizable solutions. For $m_0a>6$ there are no normalizable chiral edge states on the domain wall. If we focus just on the zero modes, we see that their number and chirality jump as a function of the parameter $m_0 a^2/R$ (or $m_0 a$ for $R=a$). The corresponding continuum limit is obtained by taking $a\rightarrow 0$ while holding $m_0 a$ constant. It naturally raises the question as to whether the Goldstone-Wilczek current also jumps as a function of $m_0 a$ to account for the current flowing on the boundary as one takes the continuum limit.
In order to understand how the Goldstone-Wilczek current compensates for the boundary current, one can integrate out the Wilson-Dirac fermion away from the domain wall as shown in \cite{Golterman:1992ub}. This leaves behind a Chern-Simons theory at low energy. The corresponding Chern-Simons coefficient is computed using a Feynman integral and can be expressed as
\beq
c=\frac{-i}{2}\frac{\epsilon_{\mu_1\mu_2\mu_3}}{3!}\int\frac{d^3p}{(2\pi)^3}\text{Tr}\{\left[S(p)\partial_{\mu_1}S(p)^{-1}\right]\left[S(p)\partial_{\mu_2}S(p)^{-1}\right]\left[S(p)\partial_{\mu_3}S(p)^{-1}\right]\}
\eeq
where $S^{-1}(p)$ is the lattice fermion propagator given by
\beq
S^{-1}(p)=\sum_{\mu=1}^d i\frac{\gamma_{\mu}\sin(p_{\mu}a)}{a}+m+r\sum_{\mu=1}^d\left[\cos(p_{\mu}a)-1\right]
\eeq
and the Chern-Simons effective action is $S_{\text{eff}}=c\,\epsilon_{\alpha_1 \alpha_2 \alpha_3}\int d^3x\, A_{\alpha_1}\partial_{\alpha_2}A_{\alpha_3}$. It was explained in \cite{Golterman:1992ub} that the Feynman integral computes the winding number of a map from a torus(momentum space) to a sphere specified by $S^{-1}(p)$. The Chern-Simons level $4\pi i c$ jumps between $1, -2$ and $1$ as $m_0 a$ is varied from $0<m_0 a<2$, $2<m_0 a<4$ and then $4<m_0 a<6$ exactly compensating for the current on the wall. These jumps in the Chern-Simons level indicate Chern insulator like topological phase transitions within the bulk away from the domain wall.  
\subsection{Continuum analysis of axion string}
\begin{figure}[h!]
\centering
\includegraphics[width=.7\textwidth]{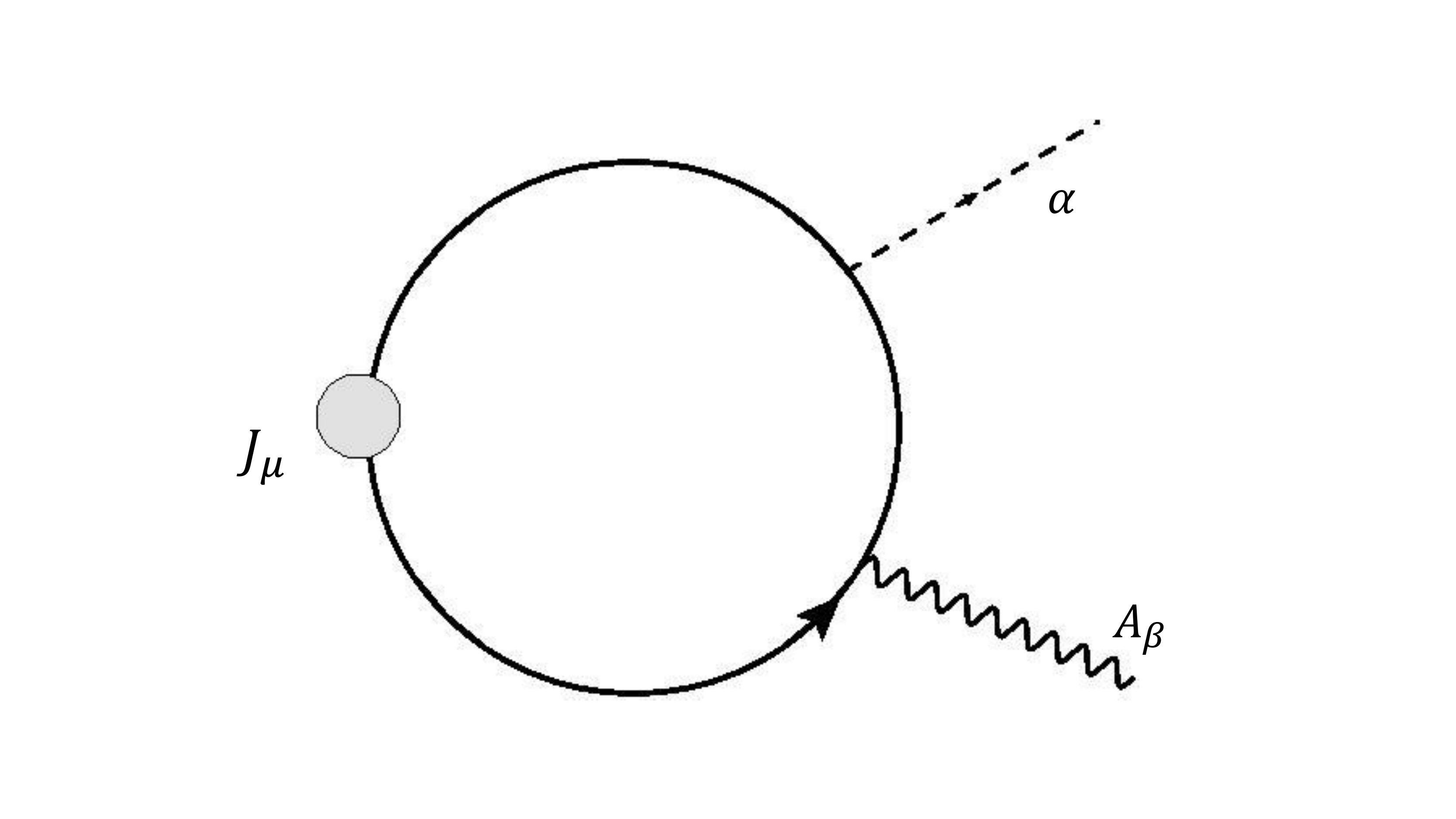}
\caption{The one-loop feynman diagram for the Goldstone-Wilczek current.}
\label{feyn2}
\end{figure}

As discussed in Callan-Harvey, besides domain wall in odd dimensional Dirac fermion theories, a Dirac fermion coupled to axion strings in even $2n+2$ dimensional theories can also exhibit chiral edge states. The continuum analysis of this was presented in Callan-Harvey which I briefly review, specializing to four dimensions, before extending the analysis to discretized space-time. 
The continuum Minkowski Lagrangian for a Dirac fermion coupled to an axion string in four space-time dimensions is given by
\beq
\mathcal{L}=\bar{\Psi}\left(i\Gamma^{\mu}\partial_{\mu}\right)\Psi-\bar{\Psi}(\phi_1-i\phi_2\bar{\Gamma})\Psi
\label{l}
\eeq
where $\bar{\Gamma}=i\Gamma^0\Gamma^1\Gamma^2\Gamma^3$. Here $\phi_1+i\phi_2\equiv \phi$ is the vacuum expectation value of a complex scalar field, the phase fluctuations of which correspond to an axion field. Since I am interested in static axion string configuration, I can take the phase  to wind by $2\pi$ around the $x^1$ axis without a loss of generality. To find the low energy spectrum on the string I can now write the equation of motion for the fermion field in the background of axion string as
\beq
\left(i\Gamma^\mu\partial_\mu\right)\Psi=\left(\phi_1-i\phi_2\bar{\Gamma}\right)\Psi.
\eeq
To look for massless chiral fermion solution to the EOM, I set $\partial_0=\partial_1=0$, and obtain
\beq
\left(i\Gamma^2\partial_2+i\Gamma^3\partial_3\right)\Psi=\left(\phi_1-i\phi_2\bar{\Gamma}\right)\Psi
\eeq
I make a specific choice for gamma matrices for convenience with $\Gamma^0=\sigma_1\otimes \sigma_1$ and $\Gamma^i=i\sigma_2\otimes\sigma_i$. Writing this equation in polar coordinates, $x^2=r\cos\theta,\,\, x^3=r\sin\theta$, I look for $\theta$ independent solutions with $\partial_{x^2}=\cos\theta\partial_r$, $\partial_{x^3}=\sin\theta\partial_r$. Thus, the EOM reduces to 
\beq
\begin{pmatrix}
0 && 0 && i\sin\theta && \cos\theta\\
0 && 0 && -\cos\theta && -i\sin\theta\\
-i\sin\theta && -\cos\theta && 0 && 0\\
\cos\theta && i\sin\theta && 0 && 0
\end{pmatrix}\partial_r\begin{pmatrix}
\Psi_+\\
\Psi_-
\end{pmatrix}=f(r)\begin{pmatrix}
e^{i\theta} && 0 && 0 && 0\\
0 && e^{i\theta} && 0 && 0\\
0 && 0 && e^{-i\theta} && 0\\
0 && 0 && 0 && e^{-i\theta}
\end{pmatrix}\begin{pmatrix}
\Psi_+\\
\Psi_-
\end{pmatrix}\nonumber\\
\label{polar}
\eeq
Here I have used $\Psi=\begin{pmatrix}
\Psi_+\\
\Psi_-
\end{pmatrix}$ and $\phi=f(r) e^{i\theta}$ where $r$ is the radial distance from the center of the string and $f(r)=|\phi |$ is only a function of the radial coordinate. The phase of the scalar field winds uniformly around the $x^1$ axis which produces the corresponding azimuthal angular dependence of $e^{i\theta}$. This equation is solved by the following ansatz 
\beq
\begin{pmatrix}0 \\ \Psi_-\end{pmatrix}&=&\eta e^{-\int_0^{\rho}f(\sigma)d\sigma}\nonumber\\
\begin{pmatrix}\Psi_+\\ 0\end{pmatrix}&=&-i\Gamma_2\begin{pmatrix}0 \\ \Psi_-\end{pmatrix}=-i\Gamma_2\eta e^{-\int_0^{\rho}f(\sigma)d\sigma}
\eeq
where $\Gamma_{\text{int}}\eta=-\eta$, $\bar{\Gamma}\eta=\eta$ and $\Gamma_\text{int}$ measures the chirality of the solution with $\Gamma_\text{int}=-\Gamma_0\Gamma_1$.
For the particular choice of gamma matrices here, one can rewrite the solution as 
\beq
\begin{pmatrix}
\Psi_+\\ \Psi_-
\end{pmatrix}=
\begin{pmatrix}
-1\\ 1\\ 1\\ 1
\end{pmatrix}e^{-\int f(\sigma)d\sigma}.
\label{sol}
\eeq
Clearly this massless chiral fermion suffers from anomaly which leads to non-conservation of fermion number current on the string in a background electric field. This is remedied by the inflowing Goldstone-Wilczek current which can be computed in perturbation theory away from the vortex core. To do this computation I parametrize $\phi_1+i\phi_2$ as $(v+\delta v)e^{i\alpha}$ where $\delta v$ and $\alpha$ vary slowly in space. I can then expand the Lagrangian in $\alpha$ and $\delta v$ to get
\beq
\mathcal{L}=\bar{\Psi}(i\Gamma^{\mu}\partial_{\mu}-v)\Psi+i\alpha v\,\bar{\Psi}\bar{\Gamma}\Psi-\delta v\bar{\Psi}\Psi+\cdots. 
\eeq
I can now attempt to compute the Goldstone Wilczek current by treating $\delta v$ and $\alpha$ as perturbation. The corresponding fermion propagator in momentum space is given by
\beq
S_F=\frac{i(\Gamma_{\mu}p^{\mu}+f)}{p^2-f^2+i\epsilon}.
\eeq
Gauging the fermion number symmetry, the fermion number current can be computed using the Feynman diagram in Fig. \ref{feyn2} as \cite{Callan:1984sa},
\beq
\mathcal{J}^{\mu}=\langle\bar{\Psi}\Gamma_{\mu}\Psi\rangle=\frac{1}{8\pi^2}\epsilon^{\mu\nu\beta\sigma}F_{\nu\beta}\partial_{\sigma}\alpha.\nonumber\\
\label{curr}
\eeq
A chiral edge state of negative chirality on the string carries a current of $-\frac{E_1}{2\pi}$ when an electric field $E_1$ is applied in the direction of the string. To see how current conservation works, I can substitute the smooth axion string (vortex) configuration with $\alpha=\theta$ where $\theta$ is the azimuthal coordinate $\theta=\tan^{-1}\left(\frac{x^3}{x^2}\right)$. For this field configuration all components of the current density in Eq. \ref{curr} is zero except the radial component. The net current flowing away from the string is then given by 
\beq
\int\mathcal{J}_r (rd\theta)=\frac{1}{2\pi}E_1
\eeq
compensating for the current flowing on the string.
  
\subsection{Crossed domain wall}
Before I discuss the lattice construction of axion string edge states, I will in this section demonstrate that an axion string configuration in the continuum can be deformed into a crossed domain wall configuration. The motivation to relate the two configurations arises from the observation that the crossed domain wall is relatively easy to discretize. As I will show, the crossed domain wall carries the same winding as the vortex configuration and hosts a chiral edge state confined to it. The latter is not surprising since the configurations carry the same winding in $\phi_1+i\phi_2$. The Goldstone-Wilczek current is only sensitive to the this winding, ensuring that the low energy fermion spectrum of the two configurations match. The crossed domain wall configuration I will consider will involve a domain wall in the field $\phi_1$ and another in $\phi_2$. Note the most general form of a crossed domain wall configuration
\beq
\phi_1&=&m_1\epsilon(x^2)+\delta m_1,\nonumber\\
\phi_2&=&m_2\epsilon(x^3)+\delta m_2\nonumber\\
\eeq
with $m_1>\delta m_1>0, m_2>\delta m_2>0$. If I define angular coordinate $\theta$ such that the four quadrants $\{x_2>0, x_3>0\}$, $\{x_2<0, x_3>0\}$, $\{x_2<0, x_3<0\}$ and $\{x_2>0, x_3<0\}$ map to $\frac{\pi}{2}>\theta>0$, $\pi>\theta>\frac{\pi}{2}$, $\frac{3\pi}{2}>\theta>\pi$ and $2\pi>\theta>\frac{3\pi}{2}$, then I can write 
\beq
\phi_1=m_1\frac{\cos\theta}{|\cos\theta|}+\delta m_1\\
\phi_2=m_2\frac{\sin\theta}{|\sin\theta|}+\delta m_2.
\eeq
I can re-parametrize $\phi_1+i\phi_2=h\left(\cos\alpha+i\sin\alpha\right)$ such that 
\beq
\tan\alpha&=&\frac{m_2\frac{\sin\theta}{|\sin\theta|}+\delta m_2}{m_1\frac{\cos\theta}{|\cos\theta|}+\delta m_1}\\
h&=&\sqrt{\left(m_1\frac{\cos\theta}{|\cos\theta|}+\delta m_1\right)^2+\left(m_2\frac{\sin\theta}{|\sin\theta|}+\delta m_2\right)^2}.
\eeq
Denoting $\alpha(\theta=0)\equiv \alpha_0$, it is easy to see that as $\theta$ is varied from $0$ to $2\pi$, $\alpha$ goes from 
$\alpha_0$ to $\alpha_0+2\pi$. Therefore $\alpha$ winds in the azimuthal direction just as one expects in a vortex. In other words
\beq
\int_0^{2\pi} \frac{1}{r}\left(\partial_{\theta}\alpha\right) r d\theta=\int_{\alpha_0}^{\alpha_0+2\pi}d\alpha=2\pi
\eeq
Of course I can choose to deform the crossed domain wall profiles slightly by replacing the step functions $\epsilon(x_2)$ and $\epsilon(x_3)$ by $\tanh(x_2)$ and $\tanh(x_3)$. This however does not affect the winding of $\alpha$. In fact, the vortex configuration used in the previous section $f(\rho)e^{i\theta}$ can simply be deformed into a crossed domain wall configuration by choosing 
$\delta m_1=\delta m_2=0$ and $m_1=m_2=f(\rho \rightarrow \infty)$. In this case we have 
$\tan(\alpha)=\frac{\tan(\theta)}{|\tan\theta|}$ and $h=f(\infty)$. 

I will now look for chiral edge state solutions confined to the crossed domain wall defect. For this I consider the equation of motion of the fermion coupled to this crossed domain wall, setting $\partial_0=\partial_1=0$
\beq
\begin{pmatrix}
0 && 0 && i\partial_3 && \partial_2\\
0 && 0 && -\partial_2 && -i\partial_3\\
-i\partial_3 && -\partial_2 && 0 && 0\\
\partial_2 && i\partial_3 && 0 && 0
\end{pmatrix}\begin{pmatrix}
\Psi_+\\
\Psi_-
\end{pmatrix}=\begin{pmatrix}
m && 0 && 0 && 0\\
0 && m && 0 && 0\\
0 && 0 && m^*&& 0\\
0 && 0 && 0 && m^*
\end{pmatrix}\begin{pmatrix}
\Psi_+\\
\Psi_-
\end{pmatrix}\nonumber\\
\eeq
where I have used $m=m_1\epsilon(x^2)+\delta m_1+i (m_2\epsilon(x^3)+\delta m_2)$. Defining $\pm m_1+\delta m_1=m_1^{\pm}$, $\pm m_2+\delta m_2=m_2^{\pm}$. It is easy to see that this equation is solved by 
\beq
\begin{pmatrix}
\Psi_+\\ \Psi_-
\end{pmatrix}=
\begin{pmatrix}
-1\\ 1\\ 1\\ 1
\end{pmatrix}\kappa(x_2, x_3)\eeq
where 
\beq
\kappa&=&
e^{-m_1^+ x^2-m_2^+ x^3}\theta(x^2)\theta(x^3)+e^{-m_1^- x^2-m_2^+ x^3}\theta(-x^2)\theta(x^3)
\nonumber\\
&&+
e^{-m_1^- x^2-m_2^- x^3}\theta(-x^2)\theta(-x^3)+e^{-m_1^+ x^2-m_2^- x^3}\theta(x^2)\theta(-x^3).
\eeq
In the limit of $\delta m_1=\delta m_2=0$ and $m_1=m_2=f_{\infty}$, the solution is 
\beq
\kappa=e^{-f_{\infty}(|x_2|+|x_3|)}.
\eeq
We therefore see that a crossed domain wall configuration carries the same winding number as in the axion string configuration. The massless edge state spectrum of the two defects are also identical, there being a chiral edge state of the same chirality confined to the core in both cases. 
\section{Discretizing space-time}
\begin{figure}[h!]
\centering
\includegraphics[width=.7\textwidth]{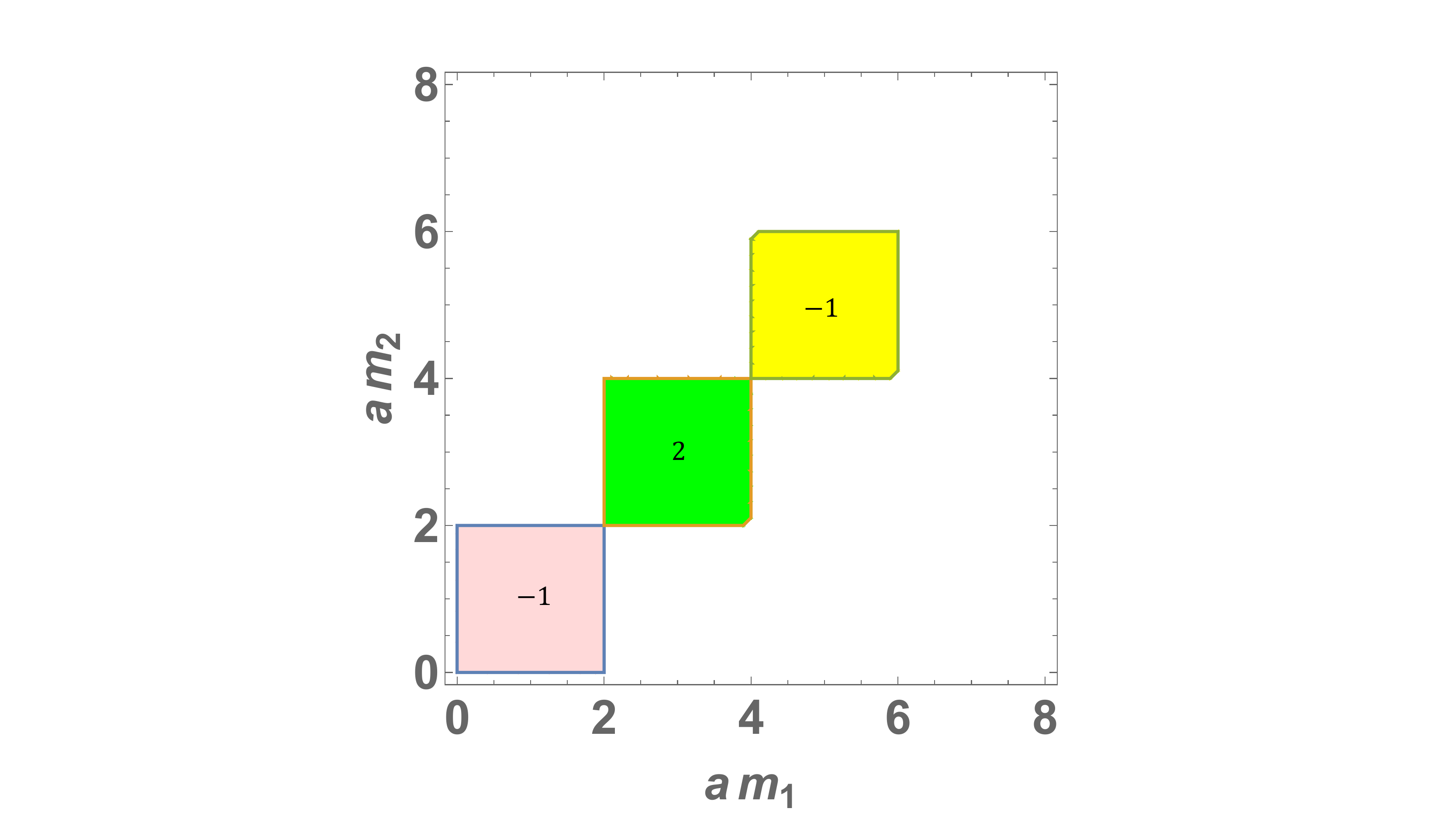}
\caption{The light red, green and yellow regions indicate the values of the parameters of the crossed domain wall, $m_1$ and $m_2$, for which there exist chiral edge states. The net chirality for the relevant parameters is also indicated inside the colored regions. In the region outside of the colored boxes, there are no chiral edge states confined to the defect.}
\label{figmake}
\end{figure}

I now consider the crossed domain wall on discrete Euclidean space-time and look for chiral edge states. I choose a square lattice such that $a^{\mu}$, the lattice spacing in $x^{\mu}$ direction is $a^{\mu}=a$. The equation of motion for the fermion in the background of a domain wall with naive discretization looks like
\beq
\frac{i\Gamma^E_{\mu}\sin(p_{\mu}a)}{a}\Psi+\left(\Gamma^E_2\nabla_2+\Gamma^E_3\nabla_3\right)\Psi=-\left(\phi_1-i\phi_2\bar{\Gamma}\right)\Psi
\eeq
where $\Gamma_\mu^E$ are Euclidean gamma matrices $\Gamma^E_0=\Gamma^0, \Gamma_i^E=-i\Gamma^i$, $\nabla_2$, $\nabla_3$ are lattice derivatives given by
\beq
\nabla_{2/3}=\frac{\delta_{x,x+a_{2/3}}-\delta_{x,x-a_{2/3}}}{2a_{2/3}}
\eeq
and the variable $\mu$ takes values $0, 1$. Note that $\bar{\Gamma}$ is defined in the text below Eq. \ref{l}. Again I have Fourier transformed the coordinates $x_0$ and $x_1$. In order to solve for massless states, I can expand the momenta around the corners of the Brillouin zone (BZ) i.e. $\{p_0a=0, p_1a=0\}$, $\{p_0a=0, p_1a=\pi\}$, $\{p_0a=\pi, p_1a=0\}$ and $\{p_0a=\pi, p_1a=\pi\}$. With an ansatz of the form
\beq
\psi=\begin{pmatrix}
-1\\ 1\\ 1\\ 1
\end{pmatrix}\varphi(p_0,p_1)\chi(x_2, x_3)
\label{ans}
\eeq
the equation for the transverse profile is
\beq
&&\nabla_2\chi=-\phi_1\chi\implies \chi(x_2+a)-\chi(x_2-a)=-2a\phi_1\chi(x_2),\nonumber\\
&&\nabla_3\chi=-\phi_2\chi\implies \chi(x_3+a)-\chi(x_3-a)=-2a\phi_2\chi(x_3).
\eeq
These two equations are solved by $\chi(x_2, x_3)=z_2^{x_2}z_3^{x_3}$ where 
\beq
z_{2/3}=\frac{-2a\phi_{1/2}-\sqrt{4a^2\phi_{1/2}^2+4}}{2}.
\eeq
So, it's clear that the conditions of normalizability for all the doublers are the same and as a result we don't have a net chirality on the string. In order to engineer net chirality on the string I will have to introduce Wilson-like terms in the Lagrangian. 
For this purpose, I propose adding to the Euclidean Lagrangian the following terms: $\frac{R}{2}\bar{\Psi}(\nabla_{2}^2+\nabla_{\text{int}}^2)\Psi-i\frac{R}{2}\bar{\Psi}\bar{\Gamma}(\nabla_{3}^2+\nabla_{\text{int}}^2)\Psi$ where $\nabla_{\text{int}}^2=\nabla_0^2+\nabla_1^2$. This shifts $\phi_{1}$ to $\phi_{1}+\frac{R}{2a^2}(\nabla_{2}^2+\nabla_{\text{int}}^2)$ and $\phi_2$ to $\phi_{2}+\frac{R}{2a^2}(\nabla_{3}^2+\nabla_{\text{int}}^2)$. Thus the equation of motions with an ansatz as in Eq. \ref{ans} is
\beq
&&\nabla_2\,\chi=-\left(\phi_1+\frac{R}{2}\nabla_2^2+\frac{R}{2a^2}\sum_{\mu}(2\cos(p_{\mu}a)-2)\right)\chi,\nonumber\\
&&\nabla_3\,\chi=-\left(\phi_2+\frac{R}{2}\nabla_3^2+\frac{R}{2a^2}\sum_{\mu}(2\cos(p_{\mu}a)-2)\right)\chi.\nonumber\\
\eeq
I set $R=a$, and solve for the transverse profile with the same ansatz as before, i.e.
\beq
\chi(x_2, x_3)=z_2^{x_2}z_3^{x_3}.
\eeq
The solutions are given by
\beq
z_2^a=-(a\phi_1-1+\sum_{\mu=0,1}(\cos(p_{\mu}a)-1))\nonumber\\
z_3^a=-(a\phi_2-1+\sum_{\mu=0,1}(\cos(p_{\mu}a)-1)).
\label{norm}
\eeq
I now impose $\delta m_1=0$ and $\delta m_2=0$ and obtain the condition of noralizability for the different doublers. If I focus on the corners of the BZ, the mode $\{p_0=0, p_1=0\}$ is normalizable for $0<m_1a<2$ and $0<m_2a<2$. The modes $\{p_0=0, p_1=\frac{\pi}{a}\}$ and $\{p_0=\frac{\pi}{a}, p_1=0\}$ are normalizable for $2<m_1a<4$ and $2<m_2a<4$. Similarly, the mode $\{p_0=\frac{\pi}{a}, p_1=\frac{\pi}{a}\}$ is normalizable for $4<m_1a<6$ and $4<m_2a<6$. There are no normalizable solutions for $m_{1/2}a>6$. I list the number and chirality of normalizable edge state solutions in table \ref{tab1} for various values of the parameters $m_1 a$ and $m_2 a$.
\begin{table}[t]
  \begin{center}
    \begin{tabular}{c|c|c|c|c|} 
 &$  2>m_1a>0 \ $&$\  4> m_1a>2 \ $&$\  6>m_1a>4\  $&$\  m_1a>6\  $\\
 \hline $2>m_2a>0 \ $&$ -1 $&$   0 $&$ \phantom{-} 0$&$ \phantom{-} 0$\\
\hline $ 4>m_2a>2  $&$  \phantom{-}0 $&$ 2 $&$ \phantom{-} 0$&$ \phantom{-} 0$\\
 \hline  $ 6>m_2a>4\ $&$  \phantom{-}0 $&$ 0 $&$  -1 $&$ \phantom{-} 0$\\
  \hline  $ m_2a>6\ $&$  \phantom{-}0 $&$ 0 $&$  \phantom{-}0 $&$ \phantom{-} 0$\\
    \end{tabular}
    \caption{Number and chirality of edge states.}
    \label{tab1}
  \end{center}
\end{table}
In Fig. \ref{figmake}, I plot the values of the parameters for which one finds chiral edge states solutions on the crossed domain wall defect.

\subsection{Goldstone-Wilczek current}
Having obtained the edge state solutions for the crossed domain wall configuration on the lattice, I will now proceed to compute the Goldstone-Wilczek current while taking the continuum limit. I will begin with the Euclidean lattice Lagnrangian
\beq
\mathcal{L}_\text{E}=\bar{\Psi}\left(\Gamma_{\mu}^E\nabla_{\mu}\right)\Psi+\bar{\Psi}\left(\phi_1+\frac{R}{2a^2}(\nabla_2^2+\nabla_{\text{int}}^2)\right)\Psi-i\bar{\Psi}\bar{\Gamma}\left(\phi_2+\frac{R}{2a^2}(\nabla_3^2+\nabla_{\text{int}}^2)\right)\Psi.
\eeq
In momentum space I can write the action as
\beq
\mathcal{S}_\text{E}&=&\int\frac{d^4p}{(2\pi)^4} \left[\bar{\Psi}(p)\left(i\Gamma_{\mu}^E\frac{\sin(ap_\mu)}{a}\right)\Psi(p)+\bar{\Psi}(p)\left(\phi_1+\frac{R}{a^2}\sum_{i=0,1,2}(\cos(p_ia)-1))\right)\Psi(p)\right.\nonumber\\
&&\left.-i\bar{\Psi}(p)\bar{\Gamma}\left(\phi_2+\frac{R}{a^2}\sum_{i=0,1,3}(\cos(p_ia)-1))\right)\Psi(p)\right].
\label{Mom1}
\eeq
Note that, $\phi_1$ and $\phi_2$ are treated as constant in space-time when going to Fourier space. This is reasonable as long as one is away from the domain walls or the domain wall profile is deformed to vary continuously across the wall. A similar condition applies to the calculation of the Goldstone-Wilczek current in the continuum where the calculation holds away from the core of the vortex.
The momentum integration in Eq. \ref{Mom1} is over one Brillouin zone (BZ). It is convenient to divide the momentum space integral around the BZ corners. Let's denote the fermion field $\Psi(p)$ near the BZ corner $p=\frac{\pi}{a}\{i, j, k, l\}$ as $\psi_{i,j,k,l}$ where $i, j, k, l$ can either be $0$ or $1$. I can now split the action over momentum integrals around the BZ corners by expanding in small $p$ as
\beq
\mathcal{S}_\text{E}&=&\int \sum_{i,j,k,l}\bar{\psi}_{i,j,k,l}(p)(i(-1)^{i}\delta_{\mu,0}\Gamma_{\mu}^E p_{\mu}+i(-1)^{j}\delta_{\mu,1}\Gamma_{\mu}^E p_{\mu}+i(-1)^{k}\delta_{\mu,2}\Gamma_{\mu}^E p_{\mu}+i(-1)^{l}\delta_{\mu,3}\Gamma_{\mu}^E p_{\mu})\psi_{i,j,k,l}(p)\nonumber\\
&+&\int\sum_{i,j,k,l}\bar{\psi}_{i,j,k,l}(p)\left[\left(\phi_1-2\frac{R}{a^2}\delta_{i,1}-2\frac{R}{a^2}\delta_{j,1}-2\frac{R}{a^2}\delta_{k,1}\right)\right.\nonumber\\
&&\hspace{2in}\left.-i\bar{\Gamma}\left(\phi_2-2\frac{R}{a^2}\delta_{i,1}-2\frac{R}{a^2}\delta_{j,1}-2\frac{R}{a^2}\delta_{l,1}\right)\right]\psi_{i,j,k,l}(p).\nonumber\\
\eeq
Thus I have rewritten the Lagrangian in terms of $16$ flavors of fermions. I can redefine the gamma matrices for these different flavors so as to absorb the factor of $(-1)^{i/j/k/l}$ in the defintion of the gamma matrices using similarity transformation. The redefined gamma matrices for the flavor $\{i,j,k,l\}$ is given by
\beq
P_{i,j,k,l}\Gamma_{0}^E P_{i,j,k,l}^{-1}=(-1)^i\Gamma_0^E,\,\,\,\,
P_{i,j,k,l}\Gamma_{1}^E P_{i,j,k,l}^{-1}&=&(-1)^j\Gamma_1^E,\,\,\, \nonumber\\
P_{i,j,k,l}\Gamma_{2}^E P_{i,j,k,l}^{-1}=(-1)^k\Gamma_2^E,\,\,\,\,
P_{i,j,k,l}\Gamma_{3}^E P_{i,j,k,l}^{-1}&=&(-1)^l\Gamma_3^E.\,\,\,\nonumber\\
\eeq 
This similarity transform takes the $\bar{\Gamma}$ matrix to
\beq
\bar{\Gamma}^{i,j,k,l}=(-1)^{i+j+k+l}\bar{\Gamma}.
\eeq
Therefore the action reduces to 
\beq
\mathcal{S}_\text{E}&=&\int \sum_{i,j,k,l}\bar{\psi}_{i,j,k,l}(p)\left[i\Gamma_{\mu}^E p_{\mu}+\left(\phi_1-2\frac{R}{a^2}\delta_{i,1}-2\frac{R}{a^2}\delta_{j,1}-2\frac{R}{a^2}\delta_{k,1}\right)\right]\psi_{i,j,k,l}(p)\nonumber\\
&-&\int\sum_{i,j,k,l}\bar{\psi}_{i,j,k,l}(p)\left[i\bar{\Gamma}^{i,j,k,l}\left(\phi_2-2\frac{R}{a^2}\delta_{i,1}-2\frac{R}{a^2}\delta_{j,1}-2\frac{R}{a^2}\delta_{l,1}\right)\right]\psi_{i,j,k,l}(p).\nonumber\\
\eeq
I want to compute the vector current for this lattice Lagrangian in the presence of a background gauge field in a crossed domain wall profile for $\phi_1$ and $\phi_2$. To do this I Fourier transform back to coordinate space. The corresponding Minkowski Lagrangian is 
\beq
&&\mathcal{L}=\sum_{i,j,k,l}\bar{\psi}_{i,j,k,l}\left[i\Gamma^{\mu}\partial_{\mu}-\left(\phi_1-2\frac{R}{a^2}\delta_{i,1}-2\frac{R}{a^2}\delta_{j,1}-2\frac{R}{a^2}\delta_{k,1}\right)\right.\nonumber\\
&&\left.\hspace{2in}+i\bar{\Gamma}^{i,j,k,l}\left(\phi_2-2\frac{R}{a^2}\delta_{i,1}-2\frac{R}{a^2}\delta_{j,1}-2\frac{R}{a^2}\delta_{l,1}\right)\right]\psi_{i,j,k,l}
\label{L2}
\eeq

For every flavor of fermion I can combine the Wilson-like terms and the fermion-scalar coupling to rewrite the action as
\beq
\mathcal{S}&=&\int \sum_{i,j,k,l}\bar{\psi}_{i,j,k,l}(p)\left(i\Gamma^{\mu}\partial_{\mu}\right)\psi_{i,j,k,l}(p)
-\int \sum_{i,j,k,l}f_{i,j,k,l}\bar{\psi}_{i,j,k,l}\psi_{i,j,k,l}\nonumber\\
&&\hspace{2in}+i\int \sum_{i,j,k,l}(-1)^{i+j+k+l}\theta_{i,j,k,l}\,f_{i,j,k,l}\,\bar{\psi}_{i,j,k,l}\bar{\Gamma}\psi_{i,j,k,l}.\nonumber\\
\eeq
where 
\beq
f_{i,j,k,l}\equiv\sqrt{\left(\phi_1-2\frac{R}{a^2}\delta_{i,1}-2\frac{R}{a^2}\delta_{j,1}-2\frac{R}{a^2}\delta_{k,1}\right)^2+\left(\phi_2-2\frac{R}{a^2}\delta_{i,1}-2\frac{R}{a^2}\delta_{j,1}-2\frac{R}{a^2}\delta_{l,1}\right)^2}\nonumber\\
\eeq
and 
\beq
\theta_{i,j,k,l}=\tan^{-1}\left((-1)^{i+j+k+l}\left(\frac{\phi_2-2\frac{R}{a^2}\delta_{i,1}-2\frac{R}{a^2}\delta_{j,1}-2\frac{R}{a^2}\delta_{l,1}}{\phi_1-2\frac{R}{a^2}\delta_{i,1}-2\frac{R}{a^2}\delta_{j,1}-2\frac{R}{a^2}\delta_{k,1}}\right)\right).
\label{theta}
\eeq
Now I set $R=a$ as before. For each flavor of fermion, I will treat $\theta_{i,j,k,l}$ as perturbation  while writing the fermion propagator as 
\beq
S_{i,j,k,l}=\frac{i(\Gamma^\mu p_{\mu}+f_{i,j,k,l})}{p^2-f_{i,j,k,l}^2+i\epsilon}.
\eeq

The vector current contribution from each flavor of fermion can be computed using perturbation theory following 
Eq. \ref{curr} using the same Feynman diagram as in Fig. \ref{feyn2}. The current for the flavor $\{i,j,k,l\}$is given by
\beq
\mathcal{J}^{i,j,k,l}_{\mu}=\frac{\epsilon_{\mu\nu\lambda\rho}}{16\pi^2}\,\partial_{\nu}\theta_{i,j,k,l}\, F^{\lambda\rho}
\eeq
which leads to the net current 
\beq
\mathcal{J}_{\mu}=\sum_{i,j,k,l}\mathcal{J}^{i,j,k,l}_{\mu}=\sum_{i,j,k,l}\frac{\epsilon_{\mu\nu\lambda\rho}}{16\pi^2}\,\partial_{\nu}\theta_{i,j,k,l}\, F^{\lambda\rho}.
\eeq
Given this expression for the current I can now verify if the integral of the divergence of this current is consistent with the number of chiral edge states on the string so as to ensure current conservation. In order to compute the net current flowing to the string from the bulk, I need to first obtain the various windings seen by the different flavors, i.e. in $\theta_{i,j,k,l}$. Note that, the overall sign of the winding will depend on the factor of $(-1)^{i+j+k+l}$ appearing in the definition $\theta_{i,j,k,l}$ in Eq. \ref{theta}. The winding in $\theta_{i,j,k,l}$ for a particular flavor can be obtained by computing
\beq
\frac{1}{2\pi}\int_0^{2\pi}\frac{1}{r}\left(\partial_{\theta}\theta_{i,j,k,l}\right)r\, d\theta
\eeq
which is always an integer. For a crossed domain wall configuration the result is a function of the domain wall heights $a|\phi_1|$ and $a|\phi_2|$. As an example let's consider $m_1=m_2=m$ and $\delta m_1=\delta m_2=0$. The corresponding windings for the different flavors are shown in table. \ref{tab2}. For $0<am<2$, a  winding of $1$ is found for the flavor $\psi_{0,0,0,0}$. The winding is zero for all the other flavors. Therefore net current flowing away from the string is
\beq
\int r d\theta\sum_{i,j,k,l}\mathcal{J}^{i,j,k,l}_{r}=\frac{E_1}{2\pi}.
\eeq
This result is consistent with there being one normalizable chiral zero mode of chirality $-1$ on the string corresponding to $\{p_0=0, p_1=0\}$ for $0<am<2$ as can be seen from Eq. \ref{norm} or table \ref{tab1}. For $2<am<4$ the flavors $\psi_{0,0,0,0}$ and $\psi_{0,0,1,1}$ see a winding of $+1$ whereas the flavors $\psi_{1,0,0,0}, \psi_{0,1,0,0},$ $\psi_{0,0,1,0}, \psi_{0,0,0,1}$ see a winding of $-1$. The rest of the flavors don't see any winding. Therefore the net current is 
\beq
\int r d\theta\sum_{i,j,k,l}\mathcal{J}^{i,j,k,l}_{r}=-\frac{E_1}{\pi}.
\eeq
This is consistent with the fact that for $2<am<4$, there are two $+1$ chirality modes on the wall, i.e. $\{p_0=0, p_1=\frac{\pi}{a}\}$ and $\{p_0=\frac{\pi}{a}, p_1=0\}$. 
Similarly, for $4<am<6$, there are seven flavors which see a positive winding of $+1$. These are $\psi_{0,0,0,0}, \psi_{0,0,1,1}, \psi_{0,1,1,0},$ $\psi_{1,1,0,0}, \psi_{1,0,0,1}, \psi_{0,1,0,1}, \psi_{1,0,1,0}$. Similarly, there are six flavors which see a winding of $-1$ and these are $\psi_{1,0,0,0}, \psi_{0,1,0,0}, \psi_{0,0,1,0}, \psi_{0,0,0,1}, \psi_{0,1,1,1}, \psi_{1,0,1,1}$. The other flavors don't see any net winding. Therefore the net current for this case is
\beq
\int r d\theta\sum_{i,j,k,l}\mathcal{J}^{i,j,k,l}_{r}=\frac{E_1}{2\pi}.
\eeq
This current is consistent with there being a single normalizable chiral zero mode of $-1$ chirality on the string, i.e. the mode $\{p_1=\frac{\pi}{a}, p_4=\frac{\pi}{a}\}$. For $am>6$, all the $16$ flavors see a winding, with eight of them seeing a winding of $+1$ and the rest $-1$. Therefore the net current is zero which is consistent with there being no normalizable edge states for $ma>6$. In table \ref{tab3} I list the Goldstone-Wilczek current as a function of the parameter $m a$ for the crossed domain wall configuration.
\begin{table}[t]
  \begin{center}
    \begin{tabular}{c|c|c|c|c|} 
 \{i, j, k, l\}&$  2>ma>0 \ $&$\  4> ma>2 \ $&$\  6>ma>4\  $&$\  ma>6\  $\\
 \hline $\{0,0,0,0\} $&$ \phantom{-}1 $&$   \phantom{-}1 $&$ \phantom{-} 1$&$ \phantom{-} 1$\\
\hline $ \{1,0,0,0\}  $&$  \phantom{-}0 $&$ -1 $&$ -1$&$ -1$\\
 \hline  $ \{0,1,0,0\}$&$  \phantom{-}0 $&$ -1 $&$  -1 $&$ -1$\\
  \hline  $\{0,0,1,0\} $&$  \phantom{-}0 $&$ -1 $&$  -1 $&$ -1$\\
   \hline $\{0,0,0,1\}$&$ \phantom{-}0 $&$   -1 $&$ -1 $&$ -1$\\
\hline $ \{1,1,0,0\}  $&$  \phantom{-}0 $&$ \phantom{-}0 $&$ \phantom{-} 1$&$ \phantom{-} 1$\\
 \hline  $\{0,1,1,0\}$&$  \phantom{-}0 $&$ \phantom{-}0 $&$  \phantom{-} 1 $&$ \phantom{-} 1$\\
  \hline  $\{0,0,1,1\} $&$  \phantom{-}0 $&$ \phantom{-}1 $&$  \phantom{-}1 $&$ \phantom{-} 1$\\
   \hline $\{1,0,1,0\} $&$ \phantom{-}0 $&$   \phantom{-}0 $&$ \phantom{-} 1$&$ \phantom{-} 1$\\
\hline $ \{0,1,0,1\}  $&$  \phantom{-}0 $&$ \phantom{-}0 $&$ \phantom{-} 1$&$ \phantom{-} 1$\\
 \hline  $ \{1,0,0,1\} $&$  \phantom{-}0 $&$ \phantom{-}0 $&$  \phantom{-} 1 $&$ \phantom{-} 1$\\
  \hline  $\{1,1,1,0\} $&$  \phantom{-}0 $&$ \phantom{-}0 $&$  \phantom{-}0 $&$ -1$\\
   \hline $\{1,1,0,1\} $&$ \phantom{-}0 $&$   \phantom{-}0 $&$ \phantom{-} 0$&$ -1$\\
\hline $\{1,0,1,1\}  $&$  \phantom{-}0 $&$ \phantom{-}0 $&$ -1$&$ -1$\\
 \hline  $\{0,1,1,1\} $&$  \phantom{-}0 $&$ \phantom{-}0 $&$  -1 $&$ -1$\\
  \hline  $\{1,1,1,1\} $&$  \phantom{-}0 $&$ \phantom{-}0 $&$  \phantom{-}0 $&$ \phantom{-} 1$\\
    \end{tabular}
    \caption{Winding in $\theta_{i,j,k,l}$ as a function of the parameter $m a$ for the $16$ different flavors. These windings are used to compute the net radial Goldstone-Wilczek current.}
    \label{tab2}
  \end{center}
\end{table}
\begin{table}[t]
  \begin{center}
    \begin{tabular}{|c|c|c|c|} 
 $  2>ma>0 \ $&$\  4> ma>2 \ $&$\  6>ma>4\  $&$\  ma>6\  $\\
 \hline $ 1 $&$   \phantom{-} -2 $&$ \phantom{-} 1$&$ \phantom{-} 0$\\
    \end{tabular}
    \caption{The table lists the net Goldstone-Wilczek current in units of $\frac{E_1}{2\pi}$. The current is obtained by taking into account the windings of different flavors as listed in table \ref{tab2}.}
    \label{tab3}
  \end{center}
\end{table}
\subsection{Crossed domain wall with unequal domain wall heights}
\label{subb}
In the previous discussion I concentrated on $m_1=m_2=m$ and $\delta m_1=\delta m_2=0$. One can repeat the analysis of the Goldstone-Wilczek current relaxing these conditions and it is easy to see that the corresponding current inflow compensates for the boundary current as required by the number and chirality of the edge modes listed in table \ref{tab1}. 

In this subsection, I will allow $m_1\neq m_2$ while holding $\delta m_1=\delta m_2=0$ and analyze the spectrum of the $16$ different flavors in the Lagrangian of Eq. \ref{L2}. I define
\beq
a\tilde{\phi}_1&=&a\phi_1-2\delta_{i,1}-2\delta_{j,1}-2\delta_{k,1},\nonumber\\
a\tilde{\phi}_2&=&a\phi_2-2\delta_{i,1}-2\delta_{j,1}-2\delta_{l,1}.\nonumber\\
\eeq
Note that $\tilde{\phi}_1$ and $\tilde{\phi}_2$ act as gaps for the different flavors and are constants away from the domain wall in $\phi_1$ and $\phi_2$. More specifically, the gap for a particular flavor is given by $\sqrt{\tilde{\phi}_1^2+\tilde{\phi}_2^2}$. It is interesting to explore the behavior of the gap and its variation in space as one changes the parameters in the theory.  
With $2>am_1>0$, the gap $a\tilde{\phi}_1$ passes through zero along $x^2=0$ domain wall, for two of the flavors $\psi_{0,0,0,0}$ and $\psi_{0,0,0,1}$. For the other $14$ flavors $\tilde{\phi}_1$ does not pass through zero anywhere in space as long as $2>a m_1>0$. If I now consider the behavior of $a\tilde{\phi}_2$, I find that for all values of $am_2$, $a\tilde{\phi}_2$ for the flavor $\psi_{0,0,0,0}$ is nonzero along the entire $x^2=0$ domain wall, except at the point $x^3=0$ . The situation for the other flavor $\psi_{0,0,0,1}$ is slightly different. For $0<am_2<2$, $a\tilde{\phi}_2$ for $\psi_{0,0,0,1}$ is nonzero in all of space. When $am_2=2$, however, $a\tilde{\phi}_2$ passes through zero, along the $x^3>0$ surface of the domain wall at $x^2=0$. Therefore the spectrum of the $\psi_{0,0,0,1}$ flavor goes gapless along a half-plane of the domain wall at $x_2=0$ when $am_2=2$. As $am_2>0$, the gap for this flavor passes through zero only at $x^2=0, x^3=0$ just as in the case of the flavor $\psi_{0,0,0,0}$. $a\tilde{\phi}_1$ and $a\tilde{\phi}_2$ does not pass through zero for any of the other flavors for $2>am_1>0$. 
As I increase $am_1$, several other flavors go gapless along the positive half of the domain wall at $x_2=0$ as $am_2$ reaches $2, 4$ and $6$. In table \ref{tab4} I show for which flavors the gap passes through zero along the entire $x^3>0$ region of the domain wall at $x^2=0$. A similar analysis can be done for the domain wall at $x^3=0$ where the gaps for the different flavors will pass through zero when $am_1$ reaches $2$, $4$ and $6$ for various values of $am_2$. Note that, the appearance of this surface along which the gap for certain flavors goes to zero coincides with the boundaries of the regions containing chiral edge states in Fig. \ref{figmake}. This indicates that the boundaries are associated with two-dimensional phase transitions.   
\begin{table}[t]
  \begin{center}
    \begin{tabular}{c|c|c|c|c|} 
 &$  2>m_1a>0 \ $&$\  4> m_1a>2 \ $&$\  6>m_1a>4\  $&$\  m_1a>6\  $\\
 \hline $m_2a=2 \ $&$ \psi_{0,0,0,1} $&$   \makecell{\psi_{1,0,0,0}, \psi_{0,1,0,0},\\ \psi_{0,0,0,1}, \psi_{0,0,1,1}} $&$ \makecell{\psi_{1,0,0,0}, \psi_{0,1,0,0}\\ \psi_{0,0,0,1},\psi_{1,0,1,0},\\ \psi_{0,1,1,0}, \psi_{0,0,1,1} }$&$ \makecell{\psi_{1,0,0,0}, \psi_{1,0,1,0}\\ \psi_{0,1,0,0}, \psi_{0,1,1,0}\\ \psi_{0,0,0,1}, \psi_{0,0,1,1}} $\\
\hline $ m_2a=4  $&$  \phantom{-}\text{\xmark} $&$ \makecell{\psi_{1,0,0,1}, \psi_{0,1,0,1}} $&$ \makecell{\psi_{1,0,0,1}, \psi_{0,1,0,1}\\ \psi_{1,0,1,1}, \psi_{0,1,1,1}\\ \psi_{1,1,0,0}}$&$ \makecell{\psi_{1,1,0,0}, \psi_{1,1,1,0}\\ \psi_{1,0,0,1}, \psi_{1,0,1,1}\\ \psi_{0,1,0,1}, \psi_{0,1,1,1}}$\\
 \hline  $ m_2a=6\ $&$  \phantom{-}\text{\xmark} $&$ \text{\xmark} $&$  \psi_{1,1,0,1} $&$ \psi_{1,1,0,1}, \psi_{1,1,1,1}$\\
    \end{tabular}
    \caption{The flavors for which the gap passes through zero along a surface in the region $x^3>0$ at $x^2=0$ ( along the domain wall) when $m_2 a=2, 4, 6$.}
    \label{tab4}
  \end{center}
\end{table}
\section{Discussion}
\subsection{Vortex configuration}
Although the crossed domain wall configuration and the vortex axion string carry the same winding and the spectrum of massless states are analogous, the details of the two systems differ. In particular, the range of the Wilson parameters and the gaps $\phi_1$ and $\phi_2$ for which one finds $1$, $2$ and $1$ chiral edge states on the defect, corresponding to the $4$ different doublers, is expected to be different for the crossed domain wall defect and the vortex. Moreover, the radial profile for the edge states in the discretized vortex configuration will be different from the radial profile obtained for the crossed domain wall configuration. The difference in the radial profile will arise from the fact that the variation in $\phi_1$ and $\phi_2$ in a vortex configuration is uniform in all of space as opposed to that in a crossed domain wall where $\phi_1$ and $\phi_2$ change rather rapidly along two surfaces while remaining constant in the rest of the bulk. 
Similarly, in a vortex configuration one will find surfaces along which the gap corresponding to different flavors will pass through zero in analogy with the discussion in section \ref{subb}. Of course, the range of the parameters for which the gap will pass through zero will be different from those obtained in table \ref{tab4}. 
\subsection{Finite volume lattice construction}
The construction of discretized axion string edge states pursued in this paper applies to infinite volume lattices. It would be interesting to implement this construction numerically which will inevitably involve a finite volume lattice. Since it would be impossible to create a single vortex of winding one in a box, one can consider a vortex and an antivortex or a vortex loop geometry so that there is no net winding in the system. The vortex and the anti vortex will host chiral edge states of opposite chirality. The number of these chiral edge states will of course depend on the ratio of the scalar vacuum expectation value to that of the Wilson parameter. This finite volume system will have a net zero chirality. The edge states will acquire a mass which is going to be exponentially suppressed in the distance between the vortex and the anti-vortex or the size of the axion loop. 

Implementing the crossed domain wall configuration in a finite volume, on the other hand will involve two vortices and two anti vortices corresponding to a wall and an anti-wall in $x_2$ and $x_3$. The crossed domain wall configuration in question are $\phi_1=m_1(\epsilon(x_2)+\epsilon(L-x_2)), \phi_2=m_2(\epsilon(x_3)+\epsilon(L-x_3))$, such that the domain wall and anti-domain walls are at a distance $L$ apart. If the defects at $\{x_2=0, x_3=0\}$ and $\{x_2=L, x_3=L\}$ carry a winding of $1$ in $\phi_1+i\phi_2$, the ones at  $\{x_2=0, x_3=L\}$ and $\{x_2=L, x_3=0\}$ carry a winding of $-1$. Then, these defects will carry chiral edge states with a net chiral imbalance provided $m_1$ and $m_2$ and the Wilson-like parameters are in the appropriate range as discussed in the paper. Locally near each of the crossed domain wall defects one will observe inflowing or outflowing current depending on the number and chirality of zero modes on the defect.
\subsection{axion insulator}
The continuum Callan-Harvey example of Dirac fermions coupled to axion field have analogs in condensed matter systems known as axion insulator. As discussed in \cite{Schuster:2016ong} different Weyl points in a Weyl semimetal can be coupled using a dynamical Higgs field $\Delta$ to reproduce the axion-fermion Lagrangian of \cite{Callan:1984sa}. When this Higgs field acquires a vacuum expectation value $\langle\Delta\rangle \neq 0$, it gaps out the Weyl points turning the Weyl semimetal into an axion insulator. In the presence of a vortex for this Higgs field where the phase of the Higgs condensate winds by $e^{i\theta}$, one obtains a string defect, i.e. axion string, which carries a chiral fermion edge state confined to it as expected from the Callan-Harvey analysis. It was shown in  \cite{Schuster:2016ong}  that the vortex configuration in the Higgs field can be engineered in an axion insulator by turning on an axial gauge field coupling to the Weyl points. 
It will be an interesting exercise to construct lattice models for axion insulators which exhibits the discrete jumps in edge states observed in this paper. The BHZ model \cite{Bernevig_2006} for quantum anomalous Hall systems may prove useful for this construction which however is beyond the scope of this paper and will be pursued as follow up work on the subject. 

\section{conclusion}
In this paper I constructed a lattice description of axion strings coupled to fermions which in the continuum is known to exhibit chiral edge states. The construction is facilitated by deforming the vortex configuration to a crossed domain wall configuration, both of which carry a winding of $1$ in the axion field. Naive discretization of space-time leads to the elimination of any net chirality on the string or the crossed domain wall core due to the presence of fermion doublers. This problem is similar to what is encountered in lattice domain wall systems where a Wilson term has to be introduced in order to engineer a net imbalance of right and left chiral edge states on the wall. Inspired by the domain wall fermion construction, I introduce Wilson-like terms in the axion string Lagrangian in order to obtain a net imbalance of right and left moving edge states on the string. As one changes the the crossed domain wall height with respect to the Wilson-like parameter, one encounters discontinuous changes in the number of chiral fermions on the string. These changes are associated with the appearance of a two dimensional surface coinciding with the domain walls where the fermion gap passes through zero indicating a two dimensional phase transition. In the presence of an electric field directed along the string, a Hall current flows from the bulk to the string. The corresponding ``Hall current" jumps between different integers, exactly compensating for the boundary current as the number and chirality of the edge states change as a function of the parameters. This current is obtained by computing a one-loop Feynman diagram which integrates out the fermion away from the defect while treating the phase variation of the axion field and the background gauge field perturbatively. There are several interesting questions that remain to be explored. One of these involves, implementing a finite volume numerical realization of both the crossed domain wall configuration and the vortex-loop configuration. It will also be interesting to explore a lattice model for the axion insulator mimicking the lattice quantum field theory described in this paper. \section{acknowledgement}
I thank Thomas Iadecola and David B. Kaplan for useful discussions. This work was supported by the Department of Energy Nuclear Physics Quantum Horizons program
through the Early Career Award DE-SC0021892.
\bibliography{axion}
\bibliographystyle{unsrt}
\end{document}